\documentclass[12pt]{article}
\usepackage{epsfig}
\usepackage{amssymb}

\begin{document}

\hfill \parbox{4cm}{UAB-FT-471\\
KA-TP-13-1999\\
hep-ph/9909397\\
September 1999}

\vskip 2cm

\begin{center}
{\Large {\sl Top quark and charged Higgs at the Tevatron-Run II}}\footnote{%
Updated talk presented in: {\em Physics at Run II Workshop on Supersymmetry/Higgs},
Fermilab, Nov. 1998, to appear in the proceedings.}

\vskip 8mm

{\large J.A. Coarasa}$^{a}${\large , Jaume Guasch}$^{b}${\large , Joan
Sol\`{a}}$^{a}$ \vskip3mm

\medskip

$^{a}${\sl Grup de F\'{\i}sica Te\`{o}rica and Institut de F\'{\i}sica d'Altes Energies,}

{\sl Universitat Aut\`{o}noma de Barcelona, E-08193, Bellaterra, Barcelona,
Catalonia, Spain}

\medskip

$^{b}${\sl Institut f\"{u}r Theorestische Physik, Universit\"{a}t Karlsruhe,}

{\sl  D-76128 Karlsruhe, Germany}

\bigskip
\end{center}

\vskip8mm\qquad \qquad \qquad \qquad \qquad

\begin{center}
{\bf ABSTRACT}
\end{center}

\begin{quotation}
\noindent We shortly review the top quark decay into charged Higgs, and
present new results on its production at the upgraded Tevatron. We have
computed the MSSM cross-section for single charged Higgs in association with the
top quark beyond the regime of on-shell $t\,\bar{t}$ production followed by
the decay $t\rightarrow H^{+}\,b,$. Our results are higher 
than recent results in the literature.
 In the case where $H^{+}$ belongs to
the Higgs sector of the MSSM, we show that the leading supersymmetric
radiative corrections may substantially increase the cross-section. Overall
we find that the charged Higgs production process can be complementary to
the neutral Higgs production processes $W\,\Phi $ and $b\,\bar{b}\,\Phi $,
which have been studied under similar circumstances. Since the neutral and
charged Higgs channels are enhanced in the same region of the parameter
space, the simultaneous detection of all these processes could be essential
for an effective experimental underpinning of the nature of these Higgs
particles at the Tevatron-Run II.
\end{quotation}

\newpage

\section{Introduction}

The top quark has been a main focus of phenomenological interest since its
discovery at the Fermilab Tevatron Collider\cite{Tevatron}. Due to its large
mass it can develop large electroweak~(EW) couplings with the Higgs bosons,
and therefore the EW quantum corrections on physical processes involving the
top quark in interplay with the Higgs sector of the model can be
substantially higher than those expected from pure gauge interactions. This
is indeed the case in some popular extensions of the Standard Model (SM) in
which the Higgs sector is enlarged, such as the Two-Higgs-Doublet Model
(2HDM) and the Minimal Supersymmetric Standard Model (MSSM)\cite{Hunter}.
Furthermore, the reach of these quantum effects has been assessed by explicit calculations of the charged
current top quark decay rates $\Gamma (t\rightarrow W^{+}\,b)$ and $\Gamma
(t\rightarrow H^{+}\,b)$\cite{tbW,CGGJS,CGHS}. And its potential impact has also been
demonstrated very recently for the FCNC top quark decay processes $\
t\rightarrow c\,\Phi $, with $\Phi =h^{0},H^{0},A^{0}$ any of the neutral
Higgs bosons in the MSSM\cite{Hunter}, by showing that the rates $\Gamma
(t\rightarrow c\,\Phi )$ --most remarkably that of the lightest CP-even
state $h^{0}$ -- can be enhanced up to reaching detection levels in future high
luminosity colliders like LHC and the LC \cite{FCNC}.

Here we wish to point out very briefly some features that could bare
relation with future Tevatron physics,  namely the top quark decay into the
charged Higgs, and new results on its production. The influence of the SUSY
quantum effects on $t\rightarrow H^{+}\,b$ can be so large on future (Run
II) Tevatron experiments that the latter could become completely distorted if
these effects are not taken into account. Moreover, the study of the
corresponding effects on the cross-section for the process $p\overline{p}%
\rightarrow t\overline{b}\,H^{-}$($\overline{t}b\,H^{+}$) is also necessary.
Especially when comparing with recent similar studies of neutral Higgs
production  $p\overline{p}\rightarrow b\overline{b}\,\Phi $ within the
MSSM\cite{Carena}. A fully-fledged analysis, however, can be quite involved as 
it might eventually require to compute at least the leading set of MSSM
radiative corrections in all these processes, neutral and charged, in order
to find out the correlation pattern among them\cite{CJS}.  In this way one
hopes to elucidate the nature (MSSM or 2HDM) of these Higgs bosons. In this
respect we recall that,  in the unconstrained 2HDM case,  there are also
several sources of large quantum effects (see Ref.\cite{CGHS}), so that it
is only after a combined study at the quantum level of the various Higgs
channels that one may be able to shed some light on whether these bosons
belong to a supersymmetric Higgs sector or not, if they are eventually
discovered.

\section{Top decay into charged Higgs}

The decay $t\rightarrow H^{+}\,b$ can be considered either in the context of
general $2$HDM's (both Type I and Type II models\cite{Hunter}) or in that of
the MSSM (whose Higgs sector is a Type II realization). In all these cases,
though for different reasons, large EW quantum effects -- on top of the
usual QCD corrections -- can play a distinguishing role. The potential large
effects stem in part from the structure of the Yukawa couplings involving
top and bottom quarks: 
\begin{equation}
\lambda _{t}\equiv {\frac{h_{t}}{g}}={\frac{m_{t}}{\sqrt{2}\,M_{W}\,\sin {%
\beta }}}\;\;\;\;\;,\;\;\;\;\;\lambda _{b}^{\left\{ I,II\right\} }\equiv {%
\frac{h_{b}}{g}}={\frac{m_{b}}{\sqrt{2}\,M_{W}\,\left\{ \sin \beta ,\cos {%
\beta }\right\} }}\,,  \label{eq:Yukawas}
\end{equation}
where the value of the parameter $\tan {\beta }$ is a most relevant one for
this kind of physics. These couplings go into the vertex $Htb$ as follows: 
\begin{equation}
{\cal L}_{Htb}={\frac{g}{\sqrt{2}M_{W}}}\,H^{+}\,\bar{t}\,[m_{t}\cot \beta
\,P_{L}+m_{b}a_{j}\,P_{R}]\,b+{\rm h.c.}\,,  \label{eq:LtbH}
\end{equation}
where $P_{L,R}=1/2(1\mp \gamma _{5})$ are the chiral projector operators,
and $a_{I}=-\cot \beta \,$, $a_{II}=+\tan {\beta }$, for Type I and Type II
respectively.

In the MSSM case, there are two overwhelming sources of leading
quantum corrections connected with bottom mass renormalization (see below). These renormalization effects are quantitatively irrelevant in the non-SUSY $2$HDM's. Alternatively, in the non-SUSY $2$HDM's the loop corrections on the vertex functions can be very large \cite{CGHS}. For, in contrast to the MSSM case, the vertex corrections do not conspire
to yield an overall negligible result because the model parameters are not
constrained by the SUSY relations.

\begin{figure}
\begin{center}
\resizebox{12cm}{!}{\includegraphics{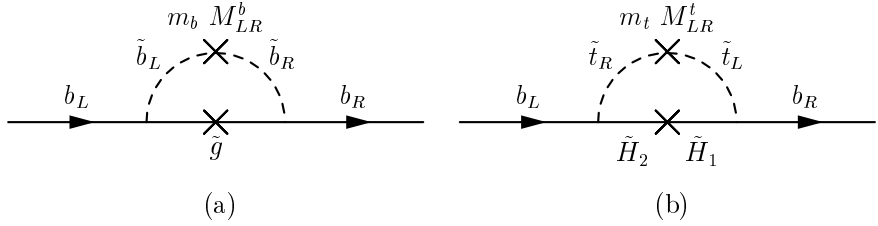}}
\end{center}
\caption{Feynman diagrams for the leading one loop  \textbf{(a)} SUSY-QCD and \textbf{(b)} SUSY-EW contributions
  to the bottom quark mass in the weak eigenstate
  basis. The crosses represent mass insertions, and $M_{LR}^{\{t,b\}}=A_{\{t,b\}}-\mu\{\cot\beta,\tan\beta\}$.
}
\end{figure}

A full one-loop calculation of $\Gamma (t\rightarrow H^{+}\,b)$ in the MSSM including
all sources of large Yukawa couplings was presented in Ref.\cite{CGGJS}.
There it was shown,  that the leading corrections were triggered by finite
threshold contributions to the SUSY mass counterterm for the bottom quark, $%
\delta m_{b}/m_{b.}$ These kind of effects can be both of SUSY-QCD and
SUSY-EW type, and originate from diagrams like the ones in Fig.\thinspace 1.
In each one of them, important contributions are singled out at high  $\tan {\beta }$  either 
because of the structure of the mixed squark propagator or by the direct
participation of the Yukawa couplings (\ref{eq:Yukawas}). Specifically,  in
the SUSY-QCD case, diagram (1a) is ``proportional'' to the gluino mass $m_{%
\widetilde{g}}$ times a supersymmetric D-term contributions $\mu \tan {\beta 
}$ (where $\mu $ is the Higgsino mass parameter), whereas in the SUSY-EW
case, diagram (1b) goes like the SUSY-breaking trilinear $A_{t}$ times the
product factor $\mu \lambda _{t}\lambda _{b}^{II}$. In both cases the
diagram increases as $\tan {\beta }$ and it would vanish in an exactly
supersymmetric world.

\begin{figure}
\begin{center}
\begin{tabular}{cc}
\resizebox{6cm}{!}{\includegraphics{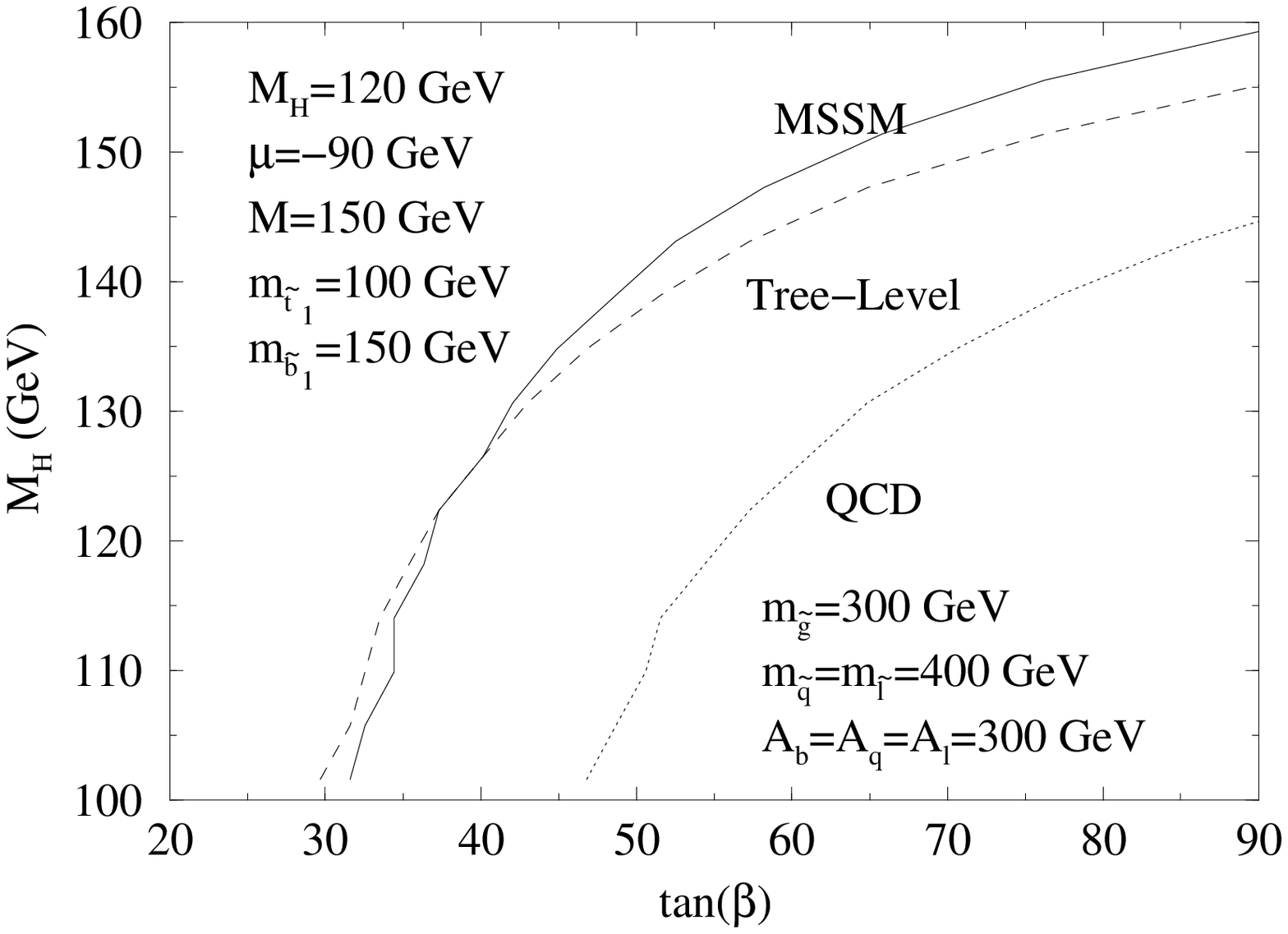}}  &
\resizebox{6cm}{!}{\includegraphics{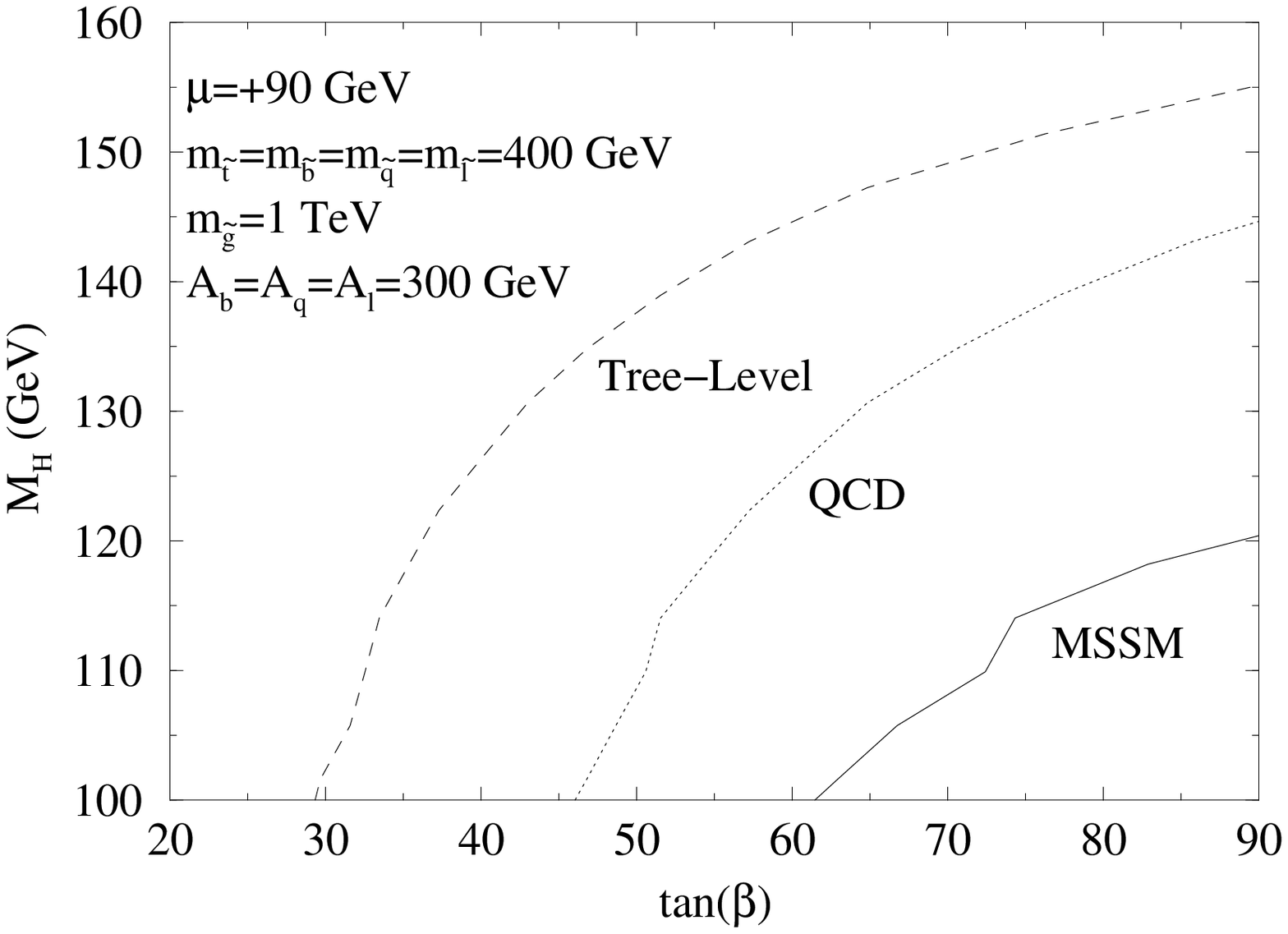}} \\
(a) &(b)
\end{tabular}
\end{center}
\caption{Exclusion regions at the 95\% C.L. in the $\tan\beta-M_H$ plane from
  the 
  $\sqrt{S}=1.8\,\,TeV$ data of the Tevatron collider, using the 
  $p\bar{p}\to t\bar{t}\to b\bar{b} \tau^{\pm} l^{\mp}
  \stackrel{\scriptscriptstyle (-)}{\nu_\tau}
  \stackrel{\scriptscriptstyle (-)}{\nu_l}$ channel. Shown are: the
  excluded region if 
  the tree-level prediction for $BR(t\to H^+b)$ is used; the excluded region
  when standard QCD (gluon) corrections are included; the excluded region when
  the total MSSM corrections are included in the prediction, for a given value
  of the MSSM parameters. The trilinear stop coupling $A_t$ is fixed by
  the restrictions imposed by the data on $b\to s\gamma$. The two figures
  correspond to the scenarios with \textbf{(a)} $\mu<0$ and \textbf{(b)}
  $\mu>0$. The excluded region is the one below each curve.}
\end{figure}

A detailed study of the implications of the SUSY quantum effects on the
Tevatron exclusion analyses of the process $t\rightarrow H^{+}\,b$ is given
in Ref.\cite{GS}. Depending on the region of the MSSM parameter space the
excluded area can be much larger (see Fig.2a) or much smaller (Fig.2b) than
expected in the absence of the SUSY effects, which amount to typical
corrections of several ten percent in the large $\tan {\beta }$ region -- $\tan
{\beta }\gtrsim 30$. 

\section{Associated production of charged Higgs and top quark at the Tevatron%
}

The most promising Tevatron process for charged Higgs production in
association with the top quark is the following: $p\overline{p}\rightarrow t%
\overline{b}\,H^{-}$($\overline{t}b\,H^{+}$). This process contributes to
the total cross-section for single top quark production\cite{Belyaev}, whose
detailed measurement is one of the main goals at the Tevatron-Run II. On the
other hand, the aforementioned charged Higgs channel is similar to the one
extensively studied for the MSSM neutral Higgs bosons in Ref.\cite{Carena},
and they both constitute leading mechanisms for charged and neutral Higgs
production at large $\tan {\beta}$. For charged Higgs masses below the
top quark mass ($M_{H}<m_{t}$), the process $p\overline{p}\rightarrow t%
\overline{b}\,H^{-}$($\overline{t}b\,H^{+}$) mainly proceeds through
factorization of $p\overline{p}\rightarrow t\overline{t}$ followed by the
decay $t\rightarrow H^{+}\,b$ and/or $\overline{t}\rightarrow H^{-}\,%
\overline{b}$. The question arises now on the behavior and quantitative
value of the cross-section for $M_{H}>m_{t}$, that is, when the process
evolves through a three-body final state diagram in which the top quark is
off-shell. The contribution from the off-shell bottom diagram is found to be
negligible.

\begin{figure}
\begin{center}
\begin{tabular}{cc}
\resizebox{8cm}{!}{\includegraphics{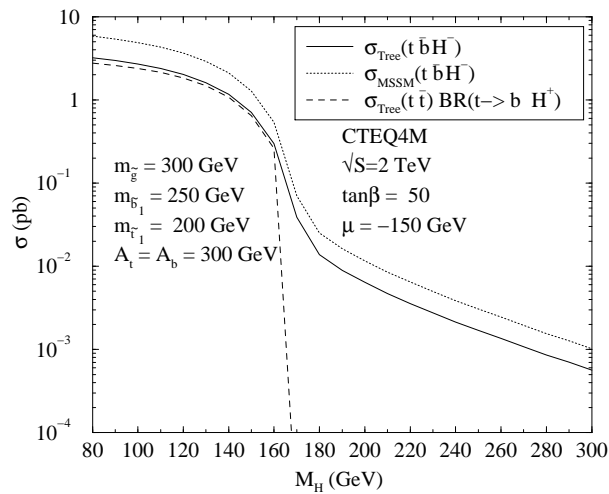}} 
\end{tabular}
\end{center}
\caption{Total cross-section for the process $p\bar{p}\to t\bar{b}H^-$ in the
  MSSM for the upgraded Tevatron ($\sqrt{S}=2\,\,TeV$) as a function of the
  charged Higgs mass. Shown are: the tree-level cross-section; the
  cross-section including the leading MSSM corrections; the tree-level
  prediction for $\sigma(p\bar{p}\to t\bar{t})\times BR(t\to H^+b)$ in the
  region where the factorization in meaningful.}
\end{figure}

In Fig. 3 we display the cross-section\footnote{See Ref.\cite{BCGS} for more
  details.} $\sigma (p\overline{p}\rightarrow t%
\overline{b}\,H^{-})$ as a function of $M_{H}$ within a range of charged
Higgs masses covering both situations $M_{H}<m_{t}$ and
$M_{H}>m_{t}$. We have fixed
$\sqrt{s}=2\,TeV$,  
corresponding to the Tevatron-Run II. It is seen that the three-body
cross-section matches the factorized cross-section in the region $M_{H}<m_{t}
$ and at the same time one sees how the cross-section extents into the
domain $M_{H}>m_{t}$ where charged Higgs bosons are not produced by real top
quark decays. Whereas the factorized process drops off very sharp near the
phase-space end, the signal for the three-body process remains significant
and could  perhaps allow to explore charged Higgs masses up to about 
$200-250\,GeV$.

In the same plot of Fig.3 we exhibit the effect of the leading SUSY
corrections for a particular choice of the MSSM parameters. We emphasize
that the leading effects have the same origin as explained in the previous
section, i.e. they stem from the $Htb$ vertex, whose leading contributions are
depicted in Fig.\,1. These effects have the remarkable property that decouple
very slowly, particularly with the gluino 
mass\cite{CGGJS,CJS}, so that for a sufficiently heavy gluino ($m_{%
\widetilde{g}}\geq 300\,GeV$) and all squark masses above $200\,GeV$, one
is guaranteed that every additional SUSY correction becomes negligible, in
particular the whole plethora of SUSY-QCD effects\cite{Zullivan} and SUSY-EW
effects\cite{Wackeroth} affecting the underlying subprocess $p\overline{p}%
\rightarrow t\overline{t}$.

The cross-section $\sigma (p\overline{p}\rightarrow t\overline{b}\,H^{-})$
can be enhanced or diminished by typical contributions that can be as large
as $\pm 50\%$ for a rather heavy SUSY spectrum involving sparticle masses of
a few hundred $GeV$ -- including gluino masses of order $1\,TeV$. Moreover,
the largest enhancements on the cross-section $\sigma (p\overline{p}%
\rightarrow t\overline{b}\,H^{-})$ occur in a region of parameter space
which is compatible with the (indirect) restrictions ($A_{t}\mu <0$) imposed
by the low-energy data on $b\rightarrow s\gamma $.

Preliminary numerical results on $\sigma (p\overline{p}\rightarrow t%
\overline{b}\,H^{-})$ were already presented in \cite{JS}. However, more
recently we have checked our calculation using the COMPHEP
package\cite{comphep}. We have included both the gluon and quark parton
distribution 
functions at the LO level, specifically the CTEQ4M\cite{cteq4m} functions
provided by COMPHEP. The factorization and renormalization scales were fixed
at the threshold value $m_{t}+M_{H}$. These settings are similar (though not
identical) to those used in Ref. \cite{Borzumati} where no SUSY radiative
effects were included. Notwithstanding, our LO cross-section (without adding
SUSY effects) is significantly larger than the one provided in that
reference. The largest deviation appears in the relevant region $M_H>m_t$ -- by roughly a factor of $2-3$. We trace the origin of the difference to the fact that we use the bottom quark pole mass, instead of the running quark mass. In this respect, we point out that the running bottom quark mass used in that reference is unusually small. Furthermore, since the QCD effects at the NLO level are expected to be large (and positive)\cite{Spira} -though they have never been explicitly computed in this particular case -- we prefer not to use the effective quark masses. In general, QCD effects on cross-sections cannot be parametrized in that way.

 To summarize, SUSY quantum effects on $p\overline{p}%
\rightarrow t\overline{b}\,H^{-}$($\overline{t}b\,H^{+}$) can be very 
important and should be taken into account in future analyses of charged
Higgs production at the Tevatron. Most important, these effects may
produce a substantial increase of the signal in regions of parameter
space which are already phenomenologically favored by other experiments. 
A more detailed analysis of the cross-section showing the
variation of the SUSY radiative corrections in different regions of the MSSM
parameter space will be presented elsewhere\cite{BCGS}.

\vskip4mm {\bf Acknowledgments.} One of the authors (JS) thanks Marcela
Carena and Carlos Wagner for discussions, and the Fermilab Theory Group for
hospitality and financial support. We thank A. Belyaev for his contribution
in the handling of the COMPHEP package. This work has also been partially
supported by CICYT under project No. AEN95-0882 and by the Deutsche
Forschungsgemeinschaft.

\end{document}